# Guiding Neutral Atoms on a Chip

N. H. Dekker,[1] C. S. Lee,[1,2] V. Lorent,[1,*] J. H. Thywissen,[1] M. Drndić,[1,2] S. P. Smith,[1] R. M. Westervelt,[1,2] and M. Prentiss[1]

[1] *Department of Physics, Harvard University, Cambridge, MA 02138, USA*

[2] *Division of Engineering and Applied Sciences, Harvard University, Cambridge, MA 02138, USA*

(August 16, 1999)

## Abstract

We demonstrate the guiding of neutral atoms by the magnetic fields due to microfabricated current-carrying wires on a chip. Atoms are guided along a magnetic field minimum parallel to and above the current-carrying wires. Two waveguide configurations are demonstrated: one using two wires with an external magnetic field, and a second using four wires without an external field. These waveguide geometries can be extended to integrated atom optics circuits, including beamsplitters.

PACS numbers: 32.60.+i, 03.75.-b, 32.80.Pj, 05.60.Gg

Typeset using REVTEX

Recent decades have seen the development of the field of integrated optics, in which separate large-scale optical elements are replaced by miniaturized devices placed together on single chips. Devices such as modulators, polarizers, couplers, and splitters have been integrated on single substrates. This has permitted rapid development in technologies such as optical communication, since in addition to a considerable cost reduction, integration has led to an increase in reliability, reproducibility, and robustness [1].

Motivated by the possibility of similar advantages for atom optics, we have demonstrated two types of magnetic waveguides that can form building blocks of integrated atom optics. The atom waveguides demonstrated in this paper are based on small, current-carrying wires on deposited on substrates using photolithography. [3]. A variety of waveguides for atoms, based both on optical [4] and magnetic interactions, have been proposed [5–8] and demonstrated [9–12]. These waveguides demonstrated in this paper are useful building block for integrated atom optics for the following reasons: (1) As the magnetic field is created by current-carrying wires, it can be dynamically reconfigured; (2) Because the wires are fabricated by photolithography, the features can be quite small ($< 150 nm$), resulting in large magnetic field gradients and curvatures [2]; (3) Although the waveguides we demonstrate in this Letter support many transverse modes, existing fabrication can produce similar devices that support just one single mode [8]; (4) The atoms follow a magnetic field zero located above the wires and are therefore easily accessible, making them compatible with atomic beamsplitters in planar wire geometries; (5) The waveguides can link other planar atom optical elements [2] fabricated on the same substrate, forming robust, complex atom optical circuits. Finally, since the interactions are magnetic, all particles that have a permanent magnetic dipole moment can be manipulated.

If the atoms follow the magnetic fields adiabatically, the interaction potential between the atoms and the magnetic fields generated by the waveguide is given by $U = \mu_B g_f m_f |B|$, where $\mu_B$ is the Bohr magneton, $g_f$ the Landé factor, $m_f$ the magnetic quantum number, and B the magnetic field. Weak-field seeking atoms feel a force towards the minimum of the magnetic field.

The two types of waveguides considered are shown in Fig. 1. A two-wire guide is constructed using two wires with oppositely-directed currents I spaced by a distance S, as shown in Fig. 1(a). The direction along which the wires are spaced is indicated by $\hat{x}$, the direction perpendicular to the plane of the wires by $\hat{y}$, and the direction parallel to the length of the wires by $\hat{z}$. The field created by the current-carrying wires is proportional to a scaling field $B_0 \hat{y} = (-2\mu_0 I)/(\pi S)\, \hat{y}$. To create a magnetic field minimum in a plane above the wires, a homogeneous external bias field $B_{ext}\hat{y}$ is applied to cancel the field due to the wires. The magnitude of the resulting magnetic total magnetic field is shown in Fig. 2(b). Both the location of the magnetic field minimum and the resulting trap depth are determined by the competition between $B_0$ and $B_{ext}$. The location of the magnetic field minimum approaches the surface as $B_{ext}$ is increased from zero until the minimum coincides with the surface at $B_{ext} = B_0$. It is desirable that the magnetic field minimum occur above the surface since atom-surface interactions can result in losses in waveguide transmission due to both adhesion and heating [13]. The maximum trap depth occurs at $B_{ext} = (1/2)B_0$, where the field minimum occurs at y $\approx$ S/2. For a fixed ratio of $B_{ext}/B_0$ it is possible to vary the trap depth without altering the location of the magnetic field minimum. Typical parameters in our experiment are S = 200 $\mu$m and I = 0.5 A, yielding $B_0$ = 20 G.

Figure 1(c) shows an alternate waveguide configuration that does not require an external field. A four-wire guide is constructed using four wires spaced by a distance S in which neighboring wires have oppositely directed currents. The inner pair produces a magnetic field proportional to a scaling field $B_{inner}\hat{y} = (-2\mu_0 I_{inner})/(\pi)\, \hat{y}$. However, instead of using $B_{ext}$ to oppose $B_{inner}$, the field of the outer pair of wires, $B_{outer}$, is used. This field scales as $B_{outer}\hat{y} = (2\mu_0 I_{outer})/(3\pi S)\, \hat{y}$. The magnitude of the resulting total magnetic field is shown in Fig. 1(d). The location of the magnetic field minimum and the trap depth are now determined by the competition between $B_{inner}$ and $B_{outer}$ [8]. Analogous to the two-wire guide, the location of the magnetic field minimum approaches the surface as $B_{outer}$ is increased, with a mimumum existing above the surface for $(1/9)\, B_{inner} < B_{outer} < B_{inner}$. Also similar to the two-wire guide, for a fixed ratio of $B_{outer}/B_{inner}$ it is possible to vary the trap depth without altering the location of the magnetic field minimum. The maximum trap depth occurs at $B_{outer} = 0.7\ B_{inner}$, where the field minimum occurs at y $\approx$ S/2..

It is interesting to consider briefly the differences between this two-wire guide and this four-wire guide. The four-wire guide does not require an external field, which may be an advantage in integrating it with other atom optical components. In addition, for small two wire atom guides (S $\sim$ 1 $\mu$m), applying an appropriate $B_{ext}$ may become difficult because the required field scales as I/S. The size of the confining potential in a four-wire guide, however, is significantly smaller than in a two-wire guide. This occurs because $B_{outer}$, which opposes $B_{inner}$, falls off with y, in contrast to $B_{ext}$, which is uniform. One may observe this difference in trap size by comparing Fig. 2(b) and Fig. 2(d) (though observe the scale change in Fig. 2(d) to accomodate four wires.)

Fabrication of waveguide wires proceeds in two steps [3]: (1) mask-based photolithography followed by liftoff and gold evaporation and (2) gold plating in solution to increase the cross-sectional area of the wires. Four parallel wires are made with a spacing of 200 $\mu$m and a wire length of 1.2 cm. The wires are formed on a sapphire substrate, chosen for its high thermal conductivity and transparency. The wires terminate in 1 mm$^2$ contact pads that support indium seals into which lead wires are pressed. Each wire has an individual feedthrough connection. This permits simple switching between two- and four-wire waveguides without changing substrates.

The sapphire substrate itself is mounted on a copper cold finger that is cooled using liquid nitrogen to a temperature of 120 K. At this temperature, the resistance of the wires decreases, permitting higher currents to be run through the wires than at room temperature; however, at low currents ($\sim$ 150 mA) we have run the waveguide at room temperature. The substrate can be rotated in the x-z plane through an angle $\theta$.

The experimental setup is shown in Fig. 2(a). Located approximately 1 cm in the $\hat{z}$-direction above the sapphire substrate is a magneto-optical trap (MOT) of Cesium atoms containing 10$^8$ atoms at a density of 5 $\cdot$ 10$^{10}$ atoms/cm$^3$. The atoms are dropped from the MOT and cooled to a temperature of approximately 15-20 $\mu$K by polarization gradient cooling. They are subsequently optically pumped to the weak-field seeking $m_f = +4$ state using an optical pump beam of waist 1 mm, $\sigma^+$ polarized, and tuned to both the F=3 $\rightarrow$ F'=4 and F=4 $\rightarrow$ F'=4 transitions, with a power of 1 $\mu$W in each component. We select these frequencies to minimize heating of the atoms.

Atoms are detected below the waveguide by monitoring the absorption of a probe beam (waist 125 $\mu$m) tuned to the F=4 $\rightarrow$ F'=5 transition on a photodiode, as shown in Fig. 2(a). To avoid integration of signal over atoms that do not interact with the waveguide, a stainless steel slit, which transmits only atoms close to the surface, is mounted approximately halfway down the length of the waveguide wires. The slit dimensions are 5.5 mm ($\hat{x}$ direction) by 0.5 mm ($\hat{y}$ direction). With the waveguide off, the small slit dimension in $\hat{y}$ limits the range of the transverse velocity of the atoms detected below the slit to approximately the recoil velocity of Cs atoms.

The time sequence of the fields is shown in Fig. 2(b). All times are relative to an initial trigger pulse. The MOT is loaded, followed by the turn-off of the gradient field at 6 ms which leads to polarization gradient cooling until 17 ms. At 33 ms, a quantization field $B_q\hat{y}$ of magnitude 0.5 G is turned on and the atoms are optically pumped. The quantization field $B_q\hat{y}$ is rotated at 65 ms into a holding field $B_h\hat{z}$ of magnitude 0.25 G. The holding field ensures that the atoms adiabatically pass from the quantization field into the much larger field of the waveguide. At 66 ms, the waveguide fields are ramped on in approximately 2 ms. The waveguide fields and holding field are ramped down at 87 ms and 88 ms, respectively. A final quantization field $B_q\hat{y}$ of magnitude 0.5 G is applied at 90 ms for detection purposes.

In this experimental setup, the following spatial signatures of guiding are predicted. The magnitude of the signal below the waveguide output should increase with the waveguide operating because for a freely expanding MOT the density of atoms above the waveguide is higher than below it, and also because the gradient of the waveguide field will draw in additional atoms transversely displaced from the input of the waveguide. If the waveguide is positioned at an angle $\theta$ with respect to the $\hat{z}$ direction defined by gravity, the position in x of the maximum waveguide output signal should be spatially separated from the position in x of the signal produced by atoms falling straight under gravity. The x position of the maximum waveguide output signal should also smoothly follow the angle $\theta$ of the waveguide.

With the pulse sequence set up as described above, we scan the probe beam below the output of a two-wire waveguide in the $\hat{x}$ direction to demonstrate these guiding effects. For the scan shown in Fig. 3, $\theta = 9.4 \pm 0.5$ degrees, the waveguide current to 0.5 A yielding $B_0 = 20$ G, and $B_{ext} = 10$ G. The signal with waveguide current off ($\triangle$) varies slowly as a function of probe position, with the peak signal at 0 mm. The signal with the waveguide on ($\bullet$ and $\square$) can be more than an order of magnitude larger, indicating that atoms have been channeled into the guide. Furthermore, its peak is spatially distinct from the free flight peak and shows a maximum at approximately -2.3 mm. The features on the signal closer to the waveguide output ($\bullet$) are slightly sharper than those farther from the waveguide output ($\square$). The blurring may be attributed to the velocity spread of the atoms leaving the waveguide. Finally, for both waveguide on traces lower signals exist to either side of the central maximum. This is consistent with atoms entering the guide experiencing a force towards the center of the guide, where the force near the center of the guide is larger than the force at the outer edge of the guide. This gradient in the force leads to depletion regions to either side of the maximum.

Besides a spatial signal of guiding, other signatures can be observed. Since only atoms in weak field seeking states are guided, if the circularity of the optical pump beam is reversed so atoms are pumped into strong field seeking states, there should be a dramatic reduction in signal; similarly, if the direction of $B_{ext}$ is reversed, the waveguide output signal should decrease. Also, if the temperature of the atoms is decreased with respect to the trap depth, the waveguide output signal should increase. Finally, as discussed above, for fixed $B_0$ (two-wire) or $B_{inner}$ (four-wire), there should be an optimum $B_{ext}$ (two-wire) or $B_{outer}$ (four-wire) that maximizes guiding.

In the experiment, we observe all of these signatures. Without the presence of an optical pump beam, the signal with the waveguide current on is typically only 20 to 30 percent larger than the signal with the waveguide current off, in contrast with Fig. 3. With the optical pump present but with polarization set to pump all atoms into the $m_f = -4$ state, no enhancement of signal is observed. Reversing the direction of $B_{ext}$ has the same effect. Varying the temperature of the Cs atoms has led to differences of a factor of five in the output of a two-wire guide. The results of varying $B_{ext}$ (two-wire) or $B_{outer}$ (four-wire) are shown in Fig. 4.

In Fig. 4(a), at low values of $B_{ext}/B_0$, both the trap depth and the magnetic field gradient are small, hence the signal is small. At higher values of $B_{ext}$, the signal increases to a maximum signal at $B_{ext}/B)0 = 0.5 \pm 0.1$. This coincides well with the value of $B_{ext}$ wich optimizes the trap depth. We observe experimentally that the value of $B_{ext}/B_0$ which maximizes the trap depth is independent of the magnetic sublevel of the atoms: without an optical pump the maximum signal as a function of $B_{ext}$ remains at $B_{ext}/B_0 = 1/2$, as expected. At higher values of $B_{ext}/B_0$, the trap depth starts to decrease and the signal correspondingly decreases. For $B_{ext} > B_0$, a magnetic field minimum no longer exists above the substrate. We believe there is still enhancement because the non-zero magnetic field gradient still results in some channeling of atoms. We have minimized this effect by pumping all the atoms into the $m_f = 4$ state; with an unpolarized source, containing atoms in lower $m_f$ states which see a weaker potential for a given current, the tail of the signal extends to 3.5 $B_{ext}/B_0$ rather than 2 $B_{ext}/B_0$.

In Fig. 4(b), a similar scan is shown for a four-wire guide. The current in the inner wires is 0.15 A, yielding $B_{inner} = 6$ G. At low values of $B_{outer}/B_{inner}$, a small signal is observed as the inner wires dominate and no magnetic field zero exists. As the value of $B_{outer}$ increases beyond 0.15 $B_{inner}$, however, a magnetic field zero is created leading to an increase in signal. We find that the signal peaks at approximately $B_{outer} = 0.26\ B_{inner}$. Beyond this point, the signal decreases as the magnetic field zero gets pushed closer to the substrate surface and the trap size decreases. We observe that the signals obtained for the four-wire guide are smaller than those for the two-wire guide. We attribute this to the lower potential depths and smaller trap size. We further observe that the peak signal does not coincide with the maximum trap depth, possibly because the peak signal also depends on trap size.

An atom beamsplitter for interferometry is an extension of these chip-based waveguides. Consider two pairs of two-wire guides that approach each other. When the distance between the pairs is large compared to their wire spacing, two separate guides exist. As the pairs are brought together, the two magnetic field zeroes approach each other while remaining above the wires. For appropriately chosen field parameters, the two magnetic field zeroes coalesce above the center of the four-wires.

In conclusion, we have demonstrated guiding of atoms with microfabricated wires on a chip. This work begins the development of atom guiding above surfaces to make integrated circuits for neutral atoms. The demonstrated waveguides represent key components of future integrated atom optics circuits.

We thank R. Younkin for comments and early work on this experiment, and M. Olshanii, M. Topinka, and G. Zabow for useful discussions. JT acknowledges support from the Fannie and John Hertz Foundation.This work was funded by ONR N00014-99-1-0347, NSF DMR-9809363, NSF PHY-9732449, and NSF PHY-9876929.

# REFERENCES

* Present address: Laboratoire de Physique des Lasers, Université Paris-Nord, F-93430 Villetaneuse, France

[1] T. Tamir, in *Guided Wave Optoelectronics*, edited by T. Tamir (Springer-Verlag, Berlin, 1988).

[2] J. D. Weinstein and K. G. Libbrecht, Phys. Rev. A **52**, 4004 (1995).

[3] M. Drndić *et al.*, Appl. Phys. Lett. **72**, 2906 (1998).

[4] J. P. Dowling and J. Gea-Banacloche, Adv. At. Mol. Opt. Phys. **37**, 1 (1996)

[5] J. A. Richmond, S. N. Chormaic, B. P. Cantwell and G. I. Opat, Acta Physica Slovaca **48**, 481 (1998).

[6] E. A. Hinds, in *New Directions in Atomic Physics*, edited by C. T. Whelan (Plenum, to be published).

[7] L. V. Hau, J. A. Golovchenko, M. M. Burns, Phys. Rev. Lett. **74**, 3138 (1995).

[8] J. H. Thywissen *et al.*, to appear Eur. Phys. J. D **7** (1999).

[9] J. Schmiedmayer, Phys. Rev. A **52**, R13 (1995); J. Schmiedmayer, Appl. Phys. B **60**, 169 (1995).

[10] J. Fortagh, A. Grossman, C. Zimmerman, and T.W. Hänsch, Phys. Rev. Lett. **81**, 5310 (1998).

[11] J. Denschlag, D. Cassettari, J. Schmiedmayer, Phys. Rev. Lett. **81**, 5310 (1999).

[12] E. Cornell (private communication).

[13] C. Henkel and M. Wilkens, Europhys. Lett. **47**, 414 (1999).

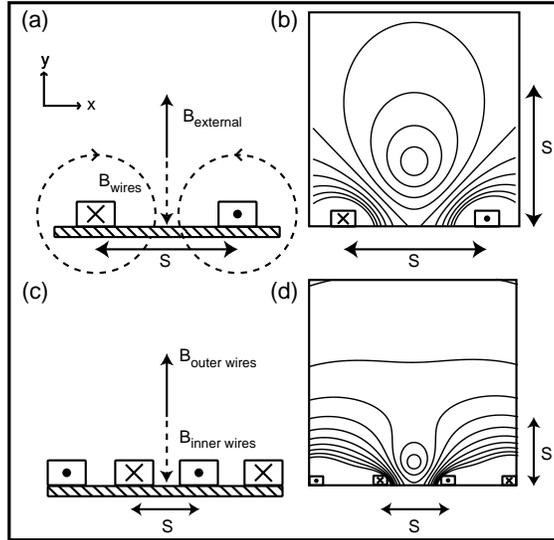

FIG. 1. (a) A two-wire magnetic g[...] G and $B_0 = 20$ G, contours of magnetic field are shown with start[...]) A four-wire magnetic guide. (d) For $B_{inner} = 20$ G and $B$ [...] alues as those in (b) are plotted.

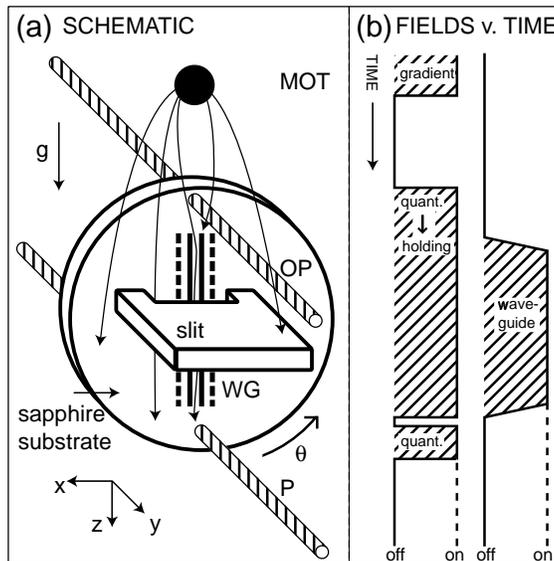

FIG. 2. (a)The experimental setup. Atoms released from the MOT fall under gravity. They pass through an optical pump beam (OP) and into the waveguide (WG). Atoms passing through a slit are detected by a probe beam P. (b) The magnetic fields as a function of time (see text.) Care is taken to ensure all fields except a quantization field are off when the probe absorption is measured.

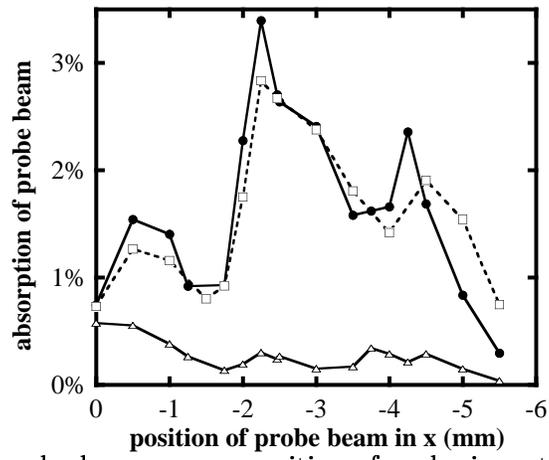

FIG. 3. Absorption of probe beam versus position of probe in x, taken with the waveguide off ($\triangle$) and the waveguide on at distances of 3.5 mm ($\bullet$) and 5.3 mm ($\square$) below the guide.

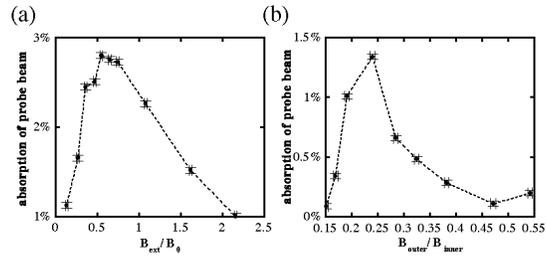

FIG. 4. (a) Absorption of probe beam below a two-wire guide as a function of $B_{ext}$. (b) Absorption of probe beam below a four-wire guide as a function of $B_{outer}$.